\documentclass[]{interact}
\usepackage{epstopdf}
\usepackage{subfigure}
\usepackage{amsthm}
\usepackage{amsmath}
\usepackage[utf8x]{inputenc}
\usepackage[justification=centering]{caption}

\usepackage{hyperref}
\usepackage{natbib}
\bibpunct[, ]{(}{)}{;}{a}{}{,}

\theoremstyle{plain}

\newtheorem{prop}{Proposition}
\theoremstyle{definition}

\theoremstyle{remark}

\begin{document}

\title{Arbitrage-free pricing of CVA for cross-currency swap with wrong-way risk under stochastic correlation modeling framework}

\author{
\name{Ashish Kumar\textsuperscript{a}\thanks{CONTACT Ashish Kumar. Email: ashish.bphu@gmail.com},  L\'aszl\'o M\'arkus  \textsuperscript{a}\thanks{CONTACT L\'aszl\'o M\'arkus  . Email: markus@cs.elte.hu} and Norbert H\'ari \thanks{CONTACT Norbert H\'ari  . Email: norbert@hari.hu}
}
\affil{\textsuperscript{a} E\"otv\"os Lor\'and University, Budapest, Hungary}
}

\maketitle
\begin{abstract}

A positive correlation between exposure and counterparty credit risk gives rise to the so-called Wrong-Way Risk (WWR). Even after a decade of the financial crisis, addressing WWR in both sound and tractable ways remains challenging. Academicians have proposed arbitrage-free set-ups through copula methods but those are computationally expensive and hard to use in practice. Resampling methods are proposed by the industry but they lack mathematical foundations. The purpose of this article is to bridge this gap between the approaches used by academicians and industry. To this end, we propose a stochastic correlation approach to asses WWR. The methods based on constant correlation to model the dependency between exposure and counterparty credit risk assume a linear dependency, thus fail to capture the tail dependence. Using a stochastic correlation we move further away from the Gaussian copula and can capture the tail risk. This effect is reflected in the results where the impact of stochastic correlation on calculated CVA is substantial when compared to the case when a high constant correlation is assumed between exposure and credit. Given the uncertainty inherent to CVA, the proposed method is believed to provide a promising way to model WWR.

\end{abstract}

\section{Introduction and Literature review}

The credit valuation adjustment (CVA) which is defined as the difference between the risk-free portfolio and the true portfolio takes into account the possibility that a counterparty might default before or at the maturity of the contract. CVA was considered negligible before the credit crisis in 2007. Roughly two-third of the credit crisis risk losses were due to CVA losses and only one-third were due to actual defaults. The amount of attention paid to counterparty risk and CVA charges has consequently increased since then.  A derivative dealer can have  multiple counterparties with over million derivatives transactions in total. CVA is calculated for each counterparty based on the total net exposure to the counterparty. Calculating CVA is computationally expensive. An excellent discussion of the issues can be found in Gregory (2009). 
While calculating CVA, it is usually assumed that the counterparty's probability of default and the dealer's exposure are independent, in  practice which usually does not hold. A situation where there is a positive dependence between the two i.e., the counterpaty default risk and its exposure  move together,  is referred to as \textit{"wrong-way risk"} (WWR). \\ 
There is no standard WWR approach widely adopted by the industry. 
There are number of models in the literature which have attempted to tackle wrong way risk problem, for example \citep{Rosen12},\citep{Ruiz14}, \citep{Hull12}. These approaches can be  broadly  classified into two groups. First family of models are those which uses copula methods to model the joint probability distribution function driving exposures values and defaults \citep{Rosen12}, typically using a Gaussian copula with a \textit{constant} correlation $\rho_{Gauss}$, and those which model the dependency between portfolio exposure and counterparty default events using an analytical approach linking portfolio value with default intensity. Both the families use change of measure. Copula based models tend to be computationally expensive because simulation of random paths from joint distributions is required. Models from the second family are faster to compute, but they require an analytical expression which links  the value of the counterparty exposure  with the default probability of the corresponding counterparty. \\
Earlier works present statistical evidence that correlation should not, indeed, be assumed
to be constant in time, see e.g \citep{Zhu99}. Furthermore, there exist several approaches which generalize the constant correlation to be stochastic, like e.g. the dynamical conditional correlation model of Engle \citep{Engle02}, local correlation models \citep{langnau09} and the Wishart auto-regressive process proposed by Gouri\'eroux et al. \citep{Gouri09}. Those mentioned papers suggest either the Jacobi process or a suitably
transformed Ornstein-Uhlenbeck process as a model for stochastic correlations. In \citep{Markus19} these stochastic correlation models are compared in terms of quantile curves of centered Kendall functions of their copulas. As correlation is not a constant quantity, the value of CVA may change with the time period. More often than not, using a linear correlation may fail to capture actual risk and thus can underestimate the value of CVA. In this paper, we revisit the problem of CVA under WWR by proposing a new method to handle it in a sound but yet tractable way. We compute CVA under WWR by using a stochastic correlation approach. This way we move further away from Gaussian copula and thus can capture tail risk, which is not the case if a constant correlation is used. The models based on constant correlation have no tail dependence, thus, fail to model the wrong-way risk in extreme events. As it is said "demon lies in the tail" and Gaussian copula based models already have been criticized enough after financial crisis. As far as authors know, no method to compute CVA by using a stochastic correlation has been proposed till date. \\ \\
This paper is organized as follows. Section 2 gives an introduction to the fundamentals of pricing CVA with WWR. Next, in section 3, we briefly describe cross-currency swap, method to compute exposure and how to model interest rate in domestic and foreign markets using change of measure technique. In section 4 we explain the modeling of stochastic default intensity using Cox, Ingersoll and Ross (CIR) model. We describe the method to model the stochastically correlated Wiener processes in section 5. Section 6 is dedicated to the introduction of our proposed model which we call as \textit{Stochastic Hybrid Model} (SHM, hereafter). In section 7 we give the description of the data used for numerical calculations. Finally, in section 8 we give an algorithmic computation of CVA under SHM and the market practiced constant correlation model and conclude in section 9.

\textbf{\section{Credit Valuation Adjustment and Wrong Way Risk}}
We denote the underlying filtered probability space as ($\Omega$, $\mathcal{F}$, $\mathcal{F}_{t}$, $\mathcal{P}$), where, $\mathcal{F}_{t}$ is the filtration which contains the market information, including default monitoring upto time t and $\mathcal{P}$ represents the physical measure. We consider two parties X and Y, where X can be an investment bank and Y which can be any financial institution entering into a cross-currency swap with a maturity of ten years. In addition to measure $\mathcal{P}$ we also suppose two risk neutral pricing measures associated with domestic and foreign markets denoted $\mathcal{Q}$ and $\mathcal{Q^{*}}$, respectively. In this paper, we only consider unilateral CVA assuming that domestic investor X is default free, therefore, all the calculations are done from the point of view of X by using domestic risk neutral measure $\mathcal{Q}$. This measure is associated with the risk-free money market account num\'eraire denoted $B_{t}$, evolving according to the risk-free rate $r_{t}$
\begin{equation}
dB_{t} = r_{t} \; B_{t} \; dt, \qquad B_{0} = 1
\end{equation}
 
Following this setup,  CVA can be computed as the $\mathcal{Q}$-expectation of the loss given default due to counterparty's default. The risk-neutral exposure faced by firm X is denoted as $V_{t}^{+}$. The general formula for the CVA can be written as 

\begin{equation}
CVA = E^{\mathcal{Q}} \; [(1-RR) \; \textbf{1}_{\tau \leq T} \; \frac{V_{\tau}^{+}}{B_{\tau}}] = (1-RR) \; E^{\mathcal{Q}} \; [E^{\mathcal{Q}} \; [\xi_{T} \; \frac{V_{\tau}^{+}}{B_{\tau}}  \;| \; \sigma(\xi_{u} \;, 0 \leq u \leq t)]] 
\end{equation}
where $E^{\mathcal{Q}}$ denotes the expectation under the risk neutral measure $\mathcal{Q}$,  $\xi = (\xi_{t})_{t\geq 0}$ is the default indicator process which is defined as $\xi_{t} = \textbf{1}_{\tau \leq t}$ and RR is the recovery rate which is assumed be constant at 25\% in this paper. Second equality follows from the assumption of constant recovery rate and the tower property. Equation (2) can be expressed in integral form as 
\begin{equation}
CVA =\; - \; (1-RR) \; \int_{0}^{T}E^{\mathcal{Q}} \; \Big[\; \frac{V_{\tau}^{+}}{B_{\tau}} \; | \; \tau \; = \; t\Big] \; dS(t),
\end{equation}
where S(t) is the risk-neutral survival probability given as 
\begin{equation}
S(t) = \; \mathcal{Q} \; [\tau \; > \; t] \;= \; E^{\mathcal{Q}} \; [\textbf{1}_{\tau \; > \; t}]
\end{equation}

Wrong-way risk (WWR) occurs when the exposure $V^{+}_t$ tends to grow as the default probability of the counterparty becomes larger. In this case, one has to account for the dependency between exposure and credit. In order to capture this effect, exposure and credit need to be modeled \textit{jointly}. The aim of this paper is to model the dependence between the exposure and default probability by using stochastic correlation in order to capture this dependence. Estimating the CVA itself is model independent, however, the underlying derivative is model dependent, this leads to more complications in the calculation of CVA.   

\textbf{\section{Exposure in Cross-Currency Swap}} 

Cross-currency swap(CCS), a special case of \textit{interest rate swap} is a derivative contract, agreed between two counterparties, which specifies the nature of the exchange of payments denominated in two different currencies. This paper takes into account the case of floating vs floating cross currency swap which is seen as an exchange of two floating rate bonds one in the domestic currency with notional $N^{d}$ and one in a foreign currency with notional $N^{f}$. The payment of floating legs  are indexed to  the domestic and  the foreign interest rates, for example LIBOR. In this paper, we assume that the domestic investor pays the floating legs in USD while the foreign investor pays in EUR, therefore, all the foreign leg cash flows need to be converted into the domestic currency by using the spot exchange rate prevailing in the market. When exchanging two currencies on the market for immediate delivery the spot exchange rate, $\psi_{0}$, is defined as

\begin{equation}
\psi_{0} = \frac{\text{Number of units of domestic currency}}{\text{Number of units of foreign currency}}
\end{equation}




The two notional amounts are related as $N_{d}$ = $\psi_{0}\;N_{f}$. In depth details on valuation of CCS can be found be in \citep{Brigo13}. 
A prototypical floating vs floating CCS exchanges cash flows between two indexed legs, starting from a future time. Consider the time partition of the interval [0,$\mathbf{T}$] as $\{0 = T_{\alpha},..., T_{\beta} = \mathbf{T}\}$, where $\mathbf{T}$ is the maturity of the CCS contract. We assume that there are 360 trading days and cash flow occur quarterly. For simplicity, we also assume that the notional is 1 unit of domestic currency and both the floating-rate payments occur at the same dates and with the same year fractions. Thus, the value of the CCS at time $t<T_{\alpha}$, from the payer's point of view  is


\begin{equation}\label{exp2}
\begin{aligned}
\Pi_{payer}(t) =   & N^{f}[\psi(t)(P^{f}(t, T_{\alpha}) - P^{f}(t, T_{\beta})] +    
\mathbf{1_{\{t=T\}}N^{f} \psi(\mathbf{T})}- \\ &  N^{d}[P(t,T_{\alpha}) - P(t, T_{\beta})]-\mathbf{1_{\{t=T\}}N^{d}} ,
\end{aligned}
\end{equation}
$\psi(t)$ is the spot foreign exchange rate. $N^{f}\psi(\mathbf{T})$ denotes the notional value received by party X at the maturity of CCS converted into domestic currency.
We can look at the value of the swap from the side of the party Y, then the value is simply obtained by changing the sign of the cash flows
\begin{equation}
\Pi_{receiver}(t) = - \Pi_{payer}(t).
\end{equation}
The detailed description on how to derive equation (\ref{exp2}) can be found in Appendix \ref{appendix:A}.\\ \\The exposure for the domestic (payer) investor can then be calculated by using the following  relation:

\begin{equation}
V^{+}(t) = max(\Pi_{payer}(t) - C(t), 0)
\end{equation}
where $\Pi_{payer}(t)$ is given in equation \eqref{exp2} and C is the collateral (if any) posted by firm Y. In the aftermath of financial crisis of 2007-08, OIS rates are usually used for discounting purposes, however, in this study we do not take into account the collateralized  portfolio (i.e., we assume firm Y posts no collateral), therefore we use LIBOR rates as proxies for risk-free rates and thus for discounting the cash flows to estimate the CVA at time 0. This reasoning can be further investigated in \citep{Hull13}.

\textbf{\subsection{Model for Multi-Currency with Short-rate Interest Rates}}

Two markets need to be modeled simultaneously, therefore, we need to understand the standard concept of no-arbitrage. Two saving accounts for the role of num\'eraire are defined, one for the domestic market
\begin{equation}
B^{d}(t) = e^{r_{d}t},
\end{equation}
and one for the foreign market 
\begin{equation}
B^{f}(t) = e^{r_{f}t}.
\end{equation}

The foreign exchange rate process $\psi(t)$ represents the domestic price at time t of one unit of the foreign currency. It is denominated in units of domestic currency per unit of the foreign currency. Foreign Exchange rate process is modeled by using the \citep{Garman83} model. This model is an extension of the Black-Scholes model in order to manage the two interest rates, one domestic and one foreign. The dynamics of $\psi(t)$ is given by the following stochastic differential equation:
\begin{equation}\label{fxx}
d\psi(t) = \psi(t) (\mu_{\psi} dt + \sigma_{\psi} dW({t})),
\end{equation}  
with constant drift $\mu_{\psi}$ and volatility $\sigma_{\psi}$ and $W(t)$ is a standard Brownian motion under the physical measure $\mathcal{P}$. 
In order to exclude arbitrage the drift of the foreign exchange rate process $\psi(t)$ need to be adjusted and its dynamics is thus given under the domestic risk neutral measure $\mathcal{Q}$. Following result is an application of Girsanov's theorem and can be found in \citep{Garman83}.

\begin{prop}\label{prop}
The dynamics of foreign exchange rate process $\psi(t)$ under the domestic martingale measure $\mathcal{Q}$ is described by
\begin{equation}
d\psi(t) = \psi(t)((r_{d} - r_{f})dt + \sigma_{\psi}dW^{\mathcal{Q}}_{\psi}(t)),
\end{equation}
\end{prop}
where $W^{\mathcal{Q}}_{\psi}(t)$ follows a Brownian motion under $\mathcal{Q}$ and $r_{d}$ and $r_{f}$ are the prevailing domestic and foreign interest rates, respectively.
Furthermore,
 
\begin{equation}
\psi(t) = \psi(0) \exp({(r_{d} - r_{f}-\frac{1}{2}\sigma^2)dt + \sigma_{\psi}W^{\mathcal{Q}}_{\psi}(t)}),
\end{equation}
Now, we are ready for the analysis of the underlying interest rate processes $r_{d}(t)$ and $r_{f}(t)$. It is assumed that the interest rate dynamics are defined via short-rate processes, which under their spot measures, i.e., $\mathcal{Q}$-domestic and $\mathcal{Q^{*}}$-foreign are driven by Hull-White one factor (HW1F) model given as,

\begin{equation}\label{domI}
dr_{d}(t) = [\theta_d(t) - \beta_{d}r_{d}(t)]dt + \sigma_{d}dW_{d}^{\mathcal{Q}}(t)
\end{equation} 

\begin{equation}\label{forI}
dr_{f}(t) = [\theta_f(t) - \beta_{f}r_{f}(t)]dt + \sigma_{f}dW_{f}^{\mathcal{Q^*}}(t),
\end{equation}
where $W_{d}^{\mathcal{Q}}(t)$ and $W_{f}^{\mathcal{Q^*}}$ are the Brownian motion under $\mathcal{Q}$ and $\mathcal{Q^{*}}$ respectively. Parameters $\beta_{d}$, $\beta_{f}$ and $\sigma_{d}$, $\sigma_{f}$ are positive constants and $\theta_d$ and $\theta_f$ are deterministic functions chosen so as to exactly fit the term structure of interest rates being currently observed in the market.\\ \\ The above processes, under the appropriate measures, are linear in their state variables, so that for a given maturity T $(0 < t < T)$ the zero-coupon bonds (ZCB) have the affine structure,
\begin{equation}\label{bpd}
P^d(t,T) =  A^d(t,T) \exp(-B^d(t,T)r_{d}(t)),
\end{equation} 
\begin{equation}\label{bpf}
P^f(t,T) =  A^f(t,T) \exp(-B^f(t,T)r_{f}(t)),
\end{equation}
where $A^d(t,T), A^f(t,T), B^d(t,T), B^f(t,T)$ are analytically known quantities and can be found in \citep{Brigo13}.\\ \\ The spot rates at time t are defined by $r_{d}(t)\equiv f_{d}^{M}(t,t)$, $r_{f}(t)\equiv f_{f}^{M}(t,t)$, where, $f_{d}^{M}(t,t)$ and $f_{f}^{M}(t,t)$ are the domestic and foreign instantaneous forward rates prevailing at time t. In general, under HW1F, the market instantaneous forward rate at time 0 for the maturity T is defined as 
\begin{equation}\label{inst}
f^{M}(0,T)  = \frac{- \partial lnP^{M}(0,T)}{\partial T}, 
\end{equation}
where $P^{M}(0,T)$ are the market implied ZCB prices prevailing at time 0 for the maturity T, we must have 

\begin{equation}
\theta(t) =  \frac{\partial f^{M}(0,T)}{\partial T} + \beta f^{M}(0,t) + \frac{\sigma^2}{2 \beta} (1 - e^{-2 \beta t}),
\end{equation}
where $\frac{\partial f^{M}}{\partial T}$ denotes partial derivative of $f^{M}$ w.r.t its second derivative. This gives us the values of $\theta_d$ and $\theta_f$ in equations (\ref{domI}) or (\ref{forI}).  \\

An alternative way to express HW1F model is by integrating equation (\ref{domI}) or (\ref{forI}), which gives us

\begin{equation}
r(t) = r(s) e^{-\beta(t-s)} + \alpha(t) - \alpha(s)e^{-\beta(t-s)} + \sigma \int_{s}^{t} e^{-\beta(t-s)} dW_{u}^{\mathcal{Q}},
\end{equation}
where,
\begin{equation}\label{alpha}
\alpha(t) = f^{M}(0,t) + \frac{\sigma^2}{2\beta^2} (1 - e^{-\beta t})^2
\end{equation}
Thus, conditional on $\mathcal{F}_{s}$,  $r_{t}$ is normally distributed under $\mathcal{Q}$ with mean and variance 
\begin{equation}\label{exd}
E[r(t)|\mathcal{F}_{s}] = r(s)e^{-\beta(t-s)} + \alpha(t) - \alpha(s) e^{-\beta(t-s)}
\end{equation}

\begin{equation}\label{vd}
Var[r({t})|\mathcal{F}_{s}] = \frac{\sigma^2}{2\beta} (1 - e^{-2\beta(t-s)})
\end{equation}

This way we no longer need to estimate $\theta_d$ and $\theta_f$ in equations (\ref{domI}),  (\ref{forI}) and the domestic and foreign interest rates can be simulated by using equations (\ref{exd}) and (\ref{vd}) which can reduce the computational complexity of the model.

\textbf{\section{Modeling Stochastic Default Intensity }}
Let $\tau$ be the stopping time that represents the default of the counterparty Y and $N({t}) = \textbf{1}_{\tau < t}$ be the right continuous increasing process  adapted to the filtration $\mathcal{F}({t}) = \sigma(N({s}), s \leq t)$ . Furthermore, $N({t})$ is a sub-martingale and $N({0})=0$.
\begin{equation}
 E[N({t})|\mathcal{F}({s})] \geq N({s}),    \qquad  \forall s<t
\end{equation}
Therefore, by Doob-Meyer decomposition, there exits a unique, increasing, predictable process    $\eta({t})$ with $\eta({0})=0$ such that $ M({t}) = N({t}) - \eta({t})$ is uniformly integrable martingale. Assuming that $\eta({t})$ is defined as 
\begin{equation}
\eta({t}) = \int_{0}^{t} \lambda({s})ds,
\end{equation}
Since $ M({t})$ is a martingale, therefore
\begin{equation}
E[M({t+dt}) - M({t})]| \mathcal{F}({t})] = 0.
\end{equation}
Thus, 
\begin{equation}
P(t < \tau < t + dt) = E[N({t+dt}) - N({t}) | \mathcal{F}({t})] = E[\int_{t}^{t+dt} \lambda({s}) |\mathcal{F}({t})]\; \approx \lambda(t)dt
\end{equation}
In this paper, $\lambda(t)$ which is called default intensity is modeled by Cox-Ingersoll-Ross (CIR) model. The risk neutral dynamics of CIR model is given as
\begin{equation}
d\lambda({t}) = \kappa_{\lambda}(\theta_{\lambda}- \lambda_{t})dt + \sigma_{\lambda} \sqrt{\lambda(t)} dW^{\mathcal{Q}}_{\lambda}(t), \qquad \kappa_{\lambda}, 
\theta_{\lambda},   \sigma_{\lambda} > 0
\end{equation}
where $W^{\mathcal{Q}}_{\lambda}(t)$ is a Wiener process under the risk neutral framework. The intensity is strictly positive if Feller condition $ 2 \kappa_{\lambda} \theta_{\lambda} > \sigma_{\lambda}^2$ is satisfied. The relationship between default intensity $\lambda_{t}$ and survival probability S(t) is given by
\begin{equation}\label{sp}
S(t) = E(e^{-\int_{0}^{t} \; \lambda(u)} \; du)=  \tilde{A}(s,t)\;e^{-\tilde{B}(s,t)\lambda_{s}}
\end{equation}
where, 
\begin{equation}
\tilde{A} = \Big[\frac{2h\;e^{(\kappa_{\lambda}+h)(t-s)/2}}{2h + (\kappa_{\lambda}+h)\{e^{h(t-s)-1}\}}\Big]^{2\kappa_{\lambda}\theta_{\lambda}/\sigma_{\lambda}^2}
\end{equation}
and
\begin{equation}
\tilde{B}(s,t) = \frac{2\{e^{h(t-s)}-1\}}{2h + (\kappa_{\lambda}+h)\{e^{h(t-s)}-1\}}
\end{equation}

\begin{equation}
h = \sqrt{\kappa_{\lambda}^2 + 2\sigma_{\lambda}^2}
\end{equation}
In many practical applications, the curve S is given exogenously from market quotes (e.g., credit default swaps).

\textbf{\section{Stochastically correlated Wiener processes}} \label{corrW}
Stochastic differential equations (SDEs) are used frequently to model data series such as interest rate, asset price, exchange rate and so on. For diffusion processes described by SDEs, the dependence between the series often originates in correlated Brownian motions driving the equations. Suppose we are given a finite time horizon $[0,T]$ and a filtered probability space $(\Omega,\mathcal{A}, \mathcal{F}_t, \mathcal{P})$ with a filtration, satisfying the usual conditions. This is e.g. the case, when the filtration is the augmented filtration of a (multidimensional) Brownian motion. Suppose, we are given two \textit{independent} Brownian motions, $W_0(t), W_1(t)$ adapted to the filtration $\mathcal{F}_t$ (i.e. the filtration is rich enough). Considering two adapted correlated Brownian motions $W_1(t)$ and $W_2(t)$ with constant correlation,  $corr( W_1(t),W_2(t))=\rho$, their quadratic covariation,
\begin{equation}\label{eqcorr}
[W_1, W_2](t)= \int\limits_0^t\rho ds=\rho \cdot t
\end{equation}
equals the covariance of $W_1(t)$, $W_2(t)$ and is symbolically written as
\begin{equation}
dW_1(t)\ dW_2(t)= \rho dt.
\end{equation}

Instead of a constant $ -1\leq\rho\leq 1$, let us given an adapted process $\rho(t)$ with $|\rho(t)|\leq 1$. Since $\rho(t)$ is bounded and the time horizon is finite $\rho(t)\in L_2\big(\Omega \times [0,T]\big)$. By using $\rho(t)$ we define the vector process 
$\mathbf{W}(t)= \binom{W_1(t)}{W_2(t)}$ as

\begin{equation}
d\mathbf{W}(t)=d\binom{W_1(t)}{W_2(t)}=
\begin{bmatrix}
1\ \ , \qquad 0\quad \ \\
\rho(t), \sqrt{1-\rho^2(t)}
\end{bmatrix} \ d\binom{W_1(t)}{W_0(t)}.
\end{equation} 

Here the second coordinate $W_2(t)$ is clearly adapted to $\mathcal{F}_t$, and as a sum of two stochastic integrals of $L_2\big(\Omega \times [0,T]\big)$ processes w.r.t. Brownian motions, it is a continuous martingale. As a direct consequence of It\^{o}'s formula the quadratic (co)variation of $\mathbf{W}(t)$ is

\begin{equation}
[\mathbf{W},\mathbf{W}](t)=\int\limits_0^t
\begin{bmatrix}
1, \rho(s) \\
\rho(s), 1
\end{bmatrix} \ dt,
\end{equation} 
and in particular, the quadratic variation of $W_2(t)$ is $t$. Being an adapted continuous martingale with quadratic variation $t$, by virtue of L\'evy's theorem \citep{Oksendal00} $W_2(t)$ is itself a one-dimensional Brownian motion. Further, the quadratic covariation of $W_1(t)$ and $W_2(t)$ is 
\begin{equation}
[W_1, W_2](t)= \int\limits_0^t\rho(s) ds.
\end{equation}
To the analogy of the constant $\rho$ in \eqref{eqcorr}, we shall call the process $\rho(t)$ the \textit{stochastic correlation} of  $W_1(t)$ and $W_2(t)$. 

\textbf{\section{Introducing Stochastic hybrid model}}\label{hyb}

The stochastic hybrid model (SHM), where we consider four processes $\psi$ the FX rate, $r_{d}$ the domestic interest rate, $r_{d}$ the foreign interest rate and $\lambda$ the default intensity, under the domestic risk neutral measure $\mathcal{Q}$ satisfy the following stochastic differential equations:

\begin{equation}\label{fx}
d\psi(t) = \psi(t)((r_{d} - r_{f})dt + \sigma_{\psi}dW^{\mathcal{Q}}_{\psi}(t)),
\end{equation}

\begin{equation}\label{ind}
dr_{d}(t) = [\theta_d(t) - \beta_{d}r_{d}(t)]dt + \sigma_{d}dW_{d}^{\mathcal{Q}}(t)
\end{equation} 

\begin{equation}\label{inf}
dr_{f}(t) = [\theta_f(t) - \beta_{f}r_{f}(t)]dt + \sigma_{f}dW_{f}^{\mathcal{Q}}(t),
\end{equation}

\begin{equation}\label{cs}
d\lambda({t}) = \kappa_{\lambda}(\theta_{\lambda}- \lambda({t}))dt + \sigma_{\lambda} \sqrt{\lambda(t)} dW_{\lambda}^{\mathcal{Q}}(t),
\end{equation}

 FX rate $\psi$ plays an important role in estimating exposure given by equation \eqref{exp2} as the foreign payments need to be converted to domestic currency. FX rate is affected by domestic and foreign interest rates, $r_{d}$ and  $r_{f}$, respectively. Therefore, if the temporal association is introduced between FX and default intensity, it's inherited by $r_{d}$ and  $r_{f}$. Two or more temporal structure between $r_{d}$ and  $r_{f}$ other than the one inherited can be introduced, however, to keep transparency and the let the dependence structure of the model readable we do not consider this approach.  

Consider the stochastic differential equations (SDE-s) for FX rate \eqref{fx} and default intensity \eqref{cs}. The two Brownian motions $dW^{\mathcal{Q}}_{\psi}(t)$ and $dW_{\lambda}^{\mathcal{Q}}(t)$ can be correlated by using the theory developed in section 5 as

\begin{equation}\label{bms}
dW_{\lambda}^{\mathcal{Q}}(t) = \rho(t)\;dW^{\mathcal{Q}}_{\psi}(t))\; + \; \sqrt{1\; - \rho^2(t)} \; dW_{0}^{\mathcal{Q}}(t), 
\end{equation}

where, the Wiener process $W_{0}^{\mathcal{Q}}(t)$ is independent of $W^{\mathcal{Q}}_{\psi}(t)$. Using L\'evy theorem (\cite{Oksendal00}) we can prove that  $W_{\lambda}^{\mathcal{Q}}(t)$ is indeed a one dimensional Wiener process.\\\\ 

\textbf{\subsection{Wiggins' Stochastic Volatility model, Discretization and Fitting}}

The underlying Brownian motions (BMs) driving the SDEs of the observed FX rate, default intensity, domestic and foreign interest rates are directly unobservable, latent processes. In order to analyze their association a suitable model is needed, its discretized version has to be fitted to the data and the underlying BMs have to be retrieved in a way that the characteristics of the BMs i.e.independence and stationarity of increments and normal distribution need to be preserved. Simple models like geometric Brownian motion are not suitable for this estimation. A large literature in the field of mathematical finance has acknowledged the temporally and stochastically changing nature of the volatility and a number of models have been created to incorporate this fact. Among them the Heston model became one of the widely adopted methodology by industry. However, in our study, when fitted to underlying data (see section 7 for the description of the data) --- the residuals obtained did not sufficiently satisfy the characteristics of the BM. This phenomenon is well-known for practitioners, and experienced by us in the current situation as well. After careful examination of a number of possibilities, in our case Wiggins' model (Wiggins (1987)) turned out to be creating probably the best residuals in terms of both normality of the distribution and the independence and stationarity of the increments.
Let us consider the following continuous time model, proposed first by Wiggins (1987),  where the spot volatility process is allowed to follow a diffusion process.
\begin{equation}
    \frac{dS(t)}{S(t)} = \mu dt + \sigma(t)dW_i(t)
    \end{equation}
    \begin{equation}
        log\left(\sigma(t)^2\right) = Y_i(t)
    \end{equation}
    \begin{equation}
        dY_i(t) = \alpha(\theta -Y(t))dt + dV(t)
    \end{equation}

    It is known that the Euler-Maruyama discretization of Wiggins' model turns into Taylor's stochastic volatility (SV) time series model with proper parametrization. The fitting of the latter via an MCMC algorithm is readily available, together with its full description, in R, in package "stochvol" \cite{Kast16}, and we use that to obtain the parameters and retrieve the residuals. The latter serve as estimations of the price driving Brownian motions, regarded as a quasi sample from its increments, and are the basis for recovering the stochastic correlation for further analysis described in the next sections. For more details on Wiggins model see \citep{Wiggins87}.\\ The powerful BDS test (after W.A.Brock, W.Dechert and J. Scheinkman) just like some more simple ones reject dependence in the residuals. Kolmogorov-Smirnov, Anderson-Darling, and Shapiro-Wilk tests reject deviation from normality of the residuals. It's well known fact that Wiener process has Gaussian and independent increments, so the generated (normal and independent)residuals can be regarded as prediction of the Wiener increments. So the cumulation of these two residuals data series lead to two Wiener processes. The "local" correlations of the two Wiener processes generated by cumulating the residuals is estimated by a \textit{sliding window technique}.

\textbf{\subsection{Interdependence Structure of the Stochastic Hybrid Model}}
In the applications the locally (in time) estimated correlation  fluctuates around a constant value in the mean reversion sense and the boundaries -1 and +1 of the correlation process are non-attractive and
unattainable. Therefore we require these properties from the model as well. The empirical stochastic correlation process $\rho(t)$ can be modeled as a transformation of a diffusion process. Teng et al \citep{Teng16} proposed the hyperbolic tangent of a mean-reverting stochastic process $G_{t}$, like an Ornstein-Uhlenbeck (OU) process 

\begin{equation}
dG(t) = \theta(\mu - G(t))dt + \sigma dW(t),
\end{equation}

\begin{equation}
G(0) = G_{0} \in \mathbb{R}
\end{equation}

For modeling stochastic correlation we map the values of OU into the [-1, 1] interval by using an hyperbolic tangent transformation y(.) as $\rho(t) = y(G(t))$. In our analysis we shall use  y(.) as the tangent hyperbolic function abbreviated in literature as "tanh" and its inverse as "atanh". To obtain the parameters of the OU model the estimated stochastic correlation process is back-transformed first and then the usual least squares estimator of the Ornstein-Uhlenbeck process is applied.  \\ \\
We suppose that the FX $\psi$, domestic interest rate $r_{d}$  and foreign interest rate $r_{f}$ follow a correlated three dimensional Wiener process. The vector process $\mathbf{W}(t)$ consisting of underlying Wiener processes of FX, domestic and foreign interest rates  is given as $\mathbf{W}(t) = \Big[W^{\mathcal{Q}}_{\psi}(t), W_{d}^{\mathcal{Q}}(t), W_{f}^{\mathcal{Q}}(t)  \Big]^{T}$, which are assumed to be correlated by using a constant correlation and given as $corr({W^{\mathcal{Q}}_{\psi}(t),W_{d}^{\mathcal{Q}}(t)}) = \rho_{12}$,  $corr({W^{\mathcal{Q}}_{\psi}(t, W_{f}^{\mathcal{Q}}(t)}) = \rho_{13}$ and $corr({W_{d}^{\mathcal{Q}}(t),W_{f}^{\mathcal{Q}}(t)}) = \rho_{23}$. We denote this correlation  matrix by $\mathcal{R}$

$$ \mathcal{R}=
\renewcommand\arraystretch{2}
\begin{bmatrix}
1 & \rho_{12} & \; \rho_{13}\\
  & 1 & \rho_{23} \\
 &   & 1 \\
\end{bmatrix} 
$$

Since the matrix $\mathcal{R}$ is positive semi-definite, it can be decomposed by Cholesky factorization 

\begin{equation}
\mathcal{R} ={A}\; {A^{T}},
\end{equation}
where A is a lower triangular matrix with zeros in the upper right corner. \\





Because the Cholesky matrix is triangular, the factors can be found by successive substitution. For the correlation matrix $\mathcal{R}$, the matrix A is given as
\[
\renewcommand\arraystretch{1.6}
\begin{bmatrix}
1 &\; 0 &\; 0 \\
\rho_{12} &\; \sqrt{1 - \rho_{12}^2} &\; 0 \\
\rho_{13} &\; \dfrac{\rho_{23} - \rho_{12}\rho_{13}}{\sqrt{1 - \rho_{12}^2}} &\; \sqrt{\dfrac{1 - \rho_{12}^2 - \rho_{13}^2 - \rho_{23}^2 + 2\rho_{12}\rho_{13}\rho_{23}} {1-\rho_{12}^2}}
\end{bmatrix}\
\]


Next, for the given vector process $\mathbf{W}(t)$ we set $d \mathbf{W}(t)$ = A\;$d \zeta$, where, $ \zeta$ is a three- dimensional vector, which is composed of independent variables all with unit variances. Therefore, with this specification complete dependence structure of the SHM model can be given as 
\begin{equation}\label{bmfx}
dW^{\mathcal{Q}}_{\psi}(t) = d\zeta_{1}
\end{equation}
\begin{equation}\label{bmind}
dW_{d}^{\mathcal{Q}}(t) = \rho_{12}\;dW^{\mathcal{Q}}_{\psi}(t)\;+\; \sqrt{1-\rho_{12}^{2}}\;d\zeta_{2}
\end{equation}
\begin{equation}\label{bminf}
dW_{f}^{\mathcal{Q}}(t) = \rho_{13}\;dW^{\mathcal{Q}}_{\psi}(t)\;+\frac{\rho_{23} - \rho_{12}\rho_{13}}{\sqrt{1 - \rho_{12}^2}}d\zeta_{2} + \;\sqrt{\frac{1 - \rho_{12}^2 - \rho_{13}^2 - \rho_{23}^2 + 2\rho_{12}\rho_{13}\rho_{23}} {1-\rho_{12}^2}}d\zeta_{3}
\end{equation}
\begin{equation}\label{bmcs}
dW_{\lambda}^{\mathcal{Q}}(t) = \rho(t)\;dW^{\mathcal{Q}}_{\psi}(t)\; + \sqrt{1-\rho(t)^{2}}\;d\zeta_{4},
\end{equation}
where \eqref{bmcs} follows from \eqref{bms} and $\zeta_{4}$ is independent of $W^{\mathcal{Q}}_{\psi}$ and $\rho(t)$ is the stochastic correlation process. By L\'evy characterisation theorem it can be showed that $W^{\mathcal{Q}}_{\psi}(t)$,  $W_{d}^{\mathcal{Q}}(t)$ and $W_{f}^{\mathcal{Q}}(t)$ are indeed Wiener processes along  with $W_{\lambda}^{\mathcal{Q}}(t)$.

\textbf{\section{Data Description}}
 
To construct the domestic and foreign yield curves we consider the USD-LIBOR, USD-SWAP and EUR-LIBOR, EUR-SWAP data observed in the market on December 23, 2018. For the implied calibration of HW1F model we use the Swaption price in USD and EUR observed in market on December 23, 2018. Other data used for the calibration purposes are implied volatility of EUR/USD FX option, Credit default swap (CDS) data of the firm Y, USD and EUR-yield curves weakly data and EUR/USD FX rate. The detailed description of the data is given in the data description table given in Appendix \ref{appendix:B}.


\textbf{\section{Monte-Carlo calculation of CVA}}

Here in this section, we give the numerical results for CVA calculation based on the methodology developed so far. When there is dependency between the default intensity of the counterparty and the exposure level, there is no simple way to calculate CVA as no closed form solution exists. We therefore use Monte carlo simulations for the results. We calculate unilateral CVA from the perspective of the firm X where we assume that recovery rate(RR) to be constant at 25\% . The credit rating of the firm Y is BBB- as per Standard \& Poor's rating agency. We assume that firm  X pays the domestic leg in Dollar  and receives the foreign leg from the firm Y in Euro. For simplicity, we assume that the notional amount is \$1 and, furthermore, the domestic leg and foreign leg payments occur at the same dates and with same year fractions . We choose time steps $t_{i}(0\leq i \leq n)$ with $t_{0} = 0$, $t_{n} = T$ and $t_{0} < t_{1} < t_{2} < ...<t_{n}$ and set 
 \begin{equation}
 CVA = (1-RR) \sum_{i=1}^{n} E^{\mathcal{Q}}[DF(t_{i})V^{+}(t_{i}) \{S(t_{i-1}) - S(t_{i})\}]
 \end{equation}

 \begin{equation}\label{cva}
   = (1-RR) \sum_{i=1}^{n} E^{\mathcal{Q}}[DF(t_{i})V^{+}(t_{i}) q(t_{i})],
 \end{equation}

where, $q(t_{i})$  is the probability of default within $(t_{i-1}, t_{i}]$. Note that $q(t_{i})$ is the unconditional risk-neutral probability of default between time $t_{i-1}$ and $t_{i}$ (as seen from time 0). It is not the probability of the default conditional on no earlier default. These $q(t_{i})'s$ can be calculated from the credit spread data of the firm Y.

In order to give a clear understanding of the method for calculating CVA under SHM model we present the steps in an algorithm representation.

\textbf{\subsection{Algorithm Implementation}}
We divide our algorithm into two parts. The first part consists of methods used for calibrating the interest rate model, stochastic default intensity model and estimating of correlation matrix $\mathcal{R}$. The second part include simulations of the SHM model.\\ \\ \\
\textbf{Step 1 - Calibration of HW1F model}
 
The first step in the implementation process is the estimation of the parameters of the underlying stochastic interest rate model i.e.,HW1F model using market implied price of financial instruments. The common method for calibration of HW1F model is by using Jamshidian's trick. Using Jamshidian's trick we can decompose the price of a swaption as a sum of Zero-Coupon bond options. For more details on Jamshidian's trick please see \citep{Jam89}. The goal is to minimise the market implied swaption price and the model implied swaption price, and therefore, to minimise the objective function 
\begin{equation}
\text{(market implied swaption price - model implied swaption price)}^2
\end{equation}
 As we have only one market implied price and two parameters including domestic and foreign markets, the mean reversion parameters $\beta_{d}$, $\beta_{f}$ and the volatility parameters $\sigma_{d}$, $\sigma_{f}$. To avoid the over-fitting, we fix the mean reversion parameters $\beta_{d}$, $\beta_{f}$  at 1\%. The estimated values of the volatility parameters of the domestic and foreign markets are (4.554\%, 3.525\%), respectively. \\ \\
 
\textbf{Step 2 - Calibration of CIR model}\\
The stochastic default intensity model is calibrated by using the historical CDS data of the firm Y. Let $s_{i}$ be the credit spread for a maturity of $t_{i}$, then the default intensity $\lambda_{i}$ can be expressed as
\begin{equation}
\lambda_{i} = \frac{s_{i}}{1-RR},
\end{equation}  
The survival probability can be calculated by using equation \eqref{fx} and therefore, default probabilities $q(t_{i})'s$ follow
\begin{equation}\label{dp}
q(t_{i}) =  S(t_{i-1}) - S(t_{i}).
\end{equation}
 
The parameters of CIR model are estimated by well-known maximum likelihood estimation.  The estimated values of the parameters $(\kappa_{\lambda}, \theta_{\lambda}, \sigma_{\lambda})$ are (0.2975, 0.3045, 0.1432), respectively. \\ \\ 
 
\textbf{Step 3 - Estimating correlation matrix $\mathcal{R}$}\\ 
Next, to fully specify the interdependence structure  our SHM model we need to estimate the correlation matrix $\mathcal{R}$. We estimate $\rho_{12}, \rho_{13}$ and $\rho_{23}$ from the historical domestic and foreign interest rates estimated from the historical yield data and the historical EUR/USD FX rate data. The residuals are estimated by fitting Wiggins model and   tested for normality and independence as explained in section 6.1. As the increments of Wiener process are independent and normal, therefore, the estimated residuals can be regarded as increments of the Wiener processes.  The estimated correlations for ($\rho_{12}, \rho_{13}$, $\rho_{23}$) are (0.1162656, 0.01965914, 0.1383345), respectively. With the estimated correlations the correlation matrix $\mathcal{R}$ can be written as 
$$
\renewcommand\arraystretch{2}
\begin{bmatrix}
1 & 0.1162656 & \; 0.01965914\\
0.1162656 & 1 & 0.1383345 \\
0.01965914& 0.1383345 & 1 \\
\end{bmatrix} 
$$
Now we ready for the simulation of the SHM model. \\

\textbf{Step 4 - Simulating SHM model and estimating exposure}\\\\
We simulate 10000 paths of the stochastic hybrid model by substituting the correlated Wiener processes from equations \eqref{bmfx}, \eqref{bmind}, \eqref{bminf} and \eqref{bmcs} into the equations \eqref{fx}, \eqref{ind}, \eqref{inf} and \eqref{cs}, respectively. Domestic and foreign interest rates are simulated by the method of exact simulation using equations \eqref{exd} and \eqref{vd}. For that cause we need to calculate $\alpha(t)$ given by equation \eqref{alpha}. The instantaneous forward rate  $f^{M}(0,t)$ is estimated by using equation \eqref{inst}. We first construct the domestic and foreign markets' yield curves by using Nelsen- Siegel model. Short ends of both the yield curves are constructed by USD and EUR LIBOR rates while the longer ends are constructed by USD and EUR swap rates. For more details on the construction of yield curve using Nelsen-Siegel model we refer reader to \citep{Nelson87}. The market implied discount factors $P^{M}(t,T)$  can be estimated from the constructed yield curves as follows 
\begin{equation}
P^{M}(t,T) = \exp(-R^{M}(t,T)\tau(t,T)),
\end{equation}

where, $R^{M}(t,T)$ is the continuously-compounded spot interest rate prevailing at time t for the maturity T in the market and $ \tau(t,T)$ is the time difference. Once we have the simulated $ r_{d}$ and $ r_{f}$, zero coupon bond prices can be simulated by equations \eqref{bpd} and \eqref{bpf}. The simulated $ r_{d}$ and $ r_{f}$ are used to simulate the FX spot rate $ \psi(t)$, given by Proposition $\eqref{prop}$. The survival and consequently default probabilities are calculated from the simulated default intensity as given by equation \eqref{sp} and \eqref{dp}, respectively. As we have assumed that the maturity of cross currency swap is 10 years and  both the domestic and foreign leg payments occur at the same dates quarterly, therefore the dimension of exposure matrix is 40 $\times$ 10000. Consequently, default probability, domestic and foreign interest rates and FX spot rates matrices also have the same dimensions. 

\begin{figure}[ht]
\centering
{%
\resizebox*{10cm}{!}{\includegraphics{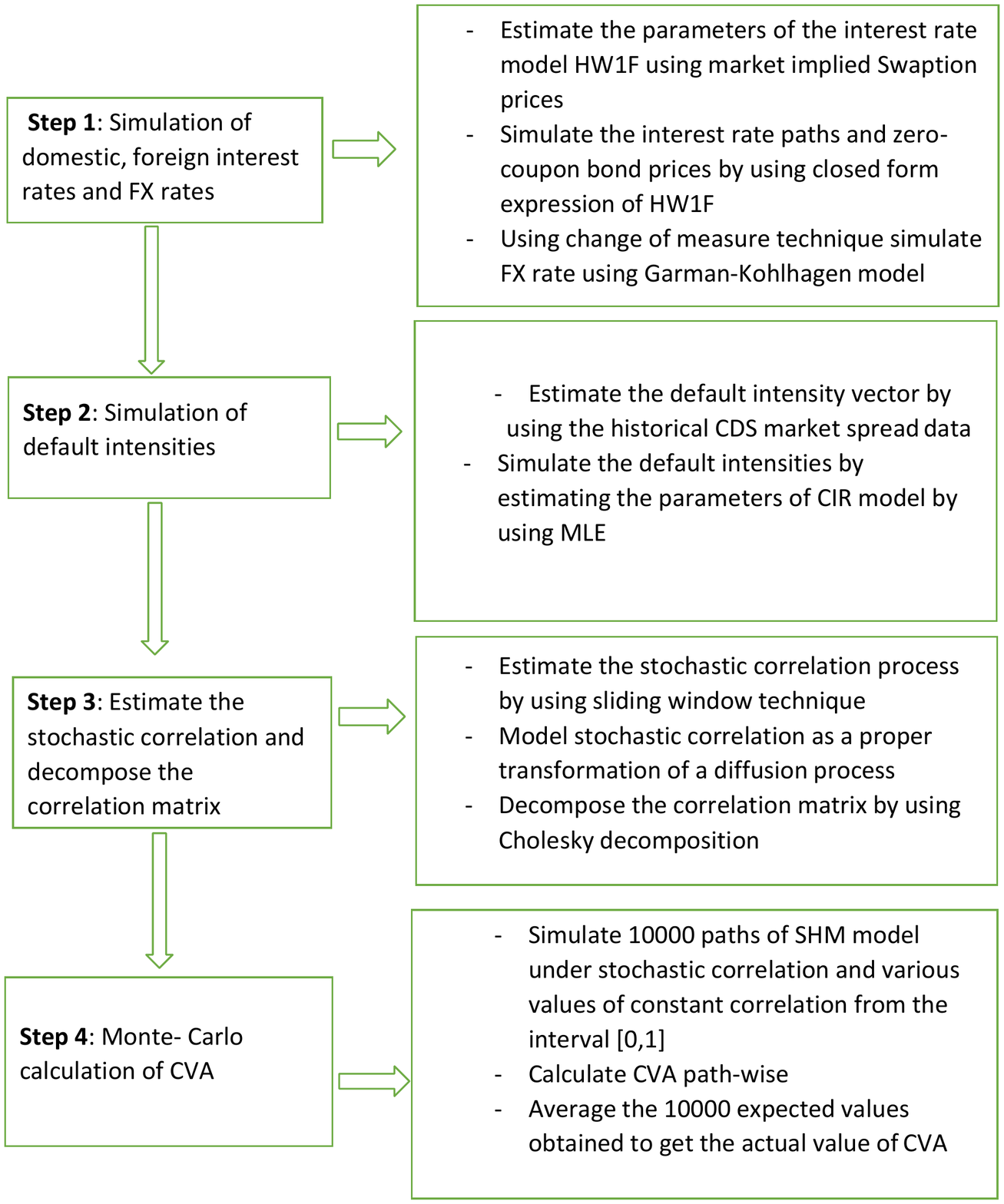}}}\hspace{10pt}
\caption{Summary of the steps for the implementation of the model.} \label{sample-figure2}
\end{figure}

\textbf{Step 5 - Calculation of CVA} \\\\
We calculate CVA using SHM model under wrong-way risk path-wise by using equation \eqref{cva}. For each simulation j, for $j \in \{1, 2, ..., 10000\}$ we estimate CVA path-wise using the discretized equation \eqref{cva}. This gives us a CVA value for every simulated path. By averaging out all 10000 pathwise CVA values we get the estimated CVA  to be charged from the counterparty. To draw a comparison between the CVA value calculated by using SHM model and the usual market practiced methods, we assume a constant correlation between default intensity and FX spot rate too, keeping the correlation matrix $\mathcal{R}$ same. The first column of the Table \ref{sample-table} corresponds to the correlation type used in SHM, keeping the correlation matrix $\mathcal{R}$ fixed. Therefore, only the correlation between FX rate and default intensity is varied to calculate CVA with wrong-way risk. In the first case, this correlation is modeled as a stochastic process as explained in section 6. In the other five cases, this correlation is kept constant and decreased linearly starting from a perfect correlation of 1 as shown in first column of Table \ref{sample-table}. Mean, standard deviation and 95\% quantile is then calculated for the 10000 estimated CVA values by generating 10000 paths of the SHM under all six models. The mean of these values corresponds the estimated CVA to be charged from the counterparty Y while entering into CCS at time 0. The steps to be followed to implement this algorithm is given in Figure \ref{sample-figure2}. The estimated CVA values under wrong-way risk using SHM and constant correlations are given in  Table \ref{sample-table}. 

\renewcommand\arraystretch{2.5}
\begin{table}[h]
\centering
\tbl{Statistics table for 10000 CVA calculated pathwise }
{\begin{tabular}{|l|l|l|l|}
\hline
$\qquad$ $\qquad$ $\quad\,$ $\big{\backslash}$ $\ $ Statistics type & CVA & Standard deviation & 95\% quantile \vspace{-0.08cm}\\ 
 Stochastic hybrid model using  $\big{\backslash}$ & & & \\
\hline
  Stochastic correlation &  0.0358243 & 0.04002753 & 0.1199763\\
\hline
$\rho$ = 1  & 0.01858226 & 0.02209931 & 0.0612432  \\
\hline
$\rho$ = 0.75 & 0.01790981 & 0.02042456 & 0.0581961 \\
 \hline
 $\rho$ = 0.50 &  0.01699416 & 0.01905533 & 0.0558105\\
\hline
 $\rho$ = 0.25  & 0.01662678 & 0.01821355 & 0.0525241 \\
\hline
 $\rho$ = 0   & 0.01577254 & 0.01696397& 0.0499159 \\
\hline
\end{tabular}}
\label{sample-table}
\end{table}

The first column shows the correlation type used between FX spot rate and default intensity in SHM, keeping the correlation matrix $\mathcal{R}$ fixed. The second column of the table contains the CVA values estimated by using Stochastic hybrid model under  different correlation type, starting from stochastic correlation. It can be seen that the CVA   increases linearly as the constant correlation between FX spot rate and default intensity increases from 0 to 1 as expected. However, this increase in the CVA values is not substantial.  CVA values estimated from  a 0 correlation to a perfect correlation between FX spot rate and default intensity increased by around 12.66\%. At the same time when CVA estimated by using stochastic correlation in SHM is 0.0358243. The CVA estimated by using stochastic correlation in SHM  almost doubles when compared to the perfect constant correlation case. To understand this impact first we need to look into the reasons why the increase in CVA values under constant correlation model is not significant. We know that correlation captures only the linear dependencies between two random variables. What if this interdependence is non-linear? More often than not, this interdependence when observed in market is usually highly non-linear.  So, while modeling wrong way risk using  constant correlation it is assumed that all the dependencies within the model are linear and, therefore, as explained in section 2, this may lead to so-called \textit{correlation risk.}  The fourth column of Table \ref{sample-table} contains the 95\% quantile values for 10000 CVA values. It is evident from these values that even if average of the  10000 simulated CVA values is 0.0358243, it can be even higher in some cases and reach upto 0.1199763 at 95\% confidence level. However, the same value for the SHM with constant correlation  is quite low when compared to SHM with stochastic correlation. This shows that using stochastic correlation the model allows more variability when compared to a constant correlation model. Other important factor why we see a higher CVA in case of stochastic correlation framework is because of the tail dependence. In general, tail dependence measures association between the extreme values of two random variables. As it is well known fact that constant correlation model does not take into account tail dependence i.e., the tail dependence under constant correlation model is zero. This is evident from the fact that Gaussian copula has tail dependence zero until the underlying correlation is perfect. Gaussian copula based credit risk models like that of \citep{Li99} have faced harsh criticism in the past, on the ground of having zero i.e. no tail dependence. Using the model with stochastic correlation we move further away from Gaussian copula and thus can be able to capture this tail risk because tail dependencies should be considered when assessing the diversification and risk of any portfolio.   

\textbf{\section{Concluding remarks}}

In this paper, we revisit the problem of pricing CVA with wrong-way risk for a cross-currency swap. Even after a decade of financial crisis, addressing wrong-way risk still remains challenging. More often than not, a constant correlation model has been used to model wrong-way risk in one way or the other which can lead to underestimation of WWR and thus the calculated value of CVA can be wrong leading to a catastrophic situations like 2007-08. Here, we propose a new approach, a stochastic correlation modeling framework to calculate CVA with wrong-way risk. This helps us to move away from the Gaussian copula models, thus, enables our model to capture the tail risk. It is well known fact that constant correlation models overlooks the tail dependence i.e., the constant correlation models have zero tail dependence until and unless a perfect positive correlation is used which may not be practically true in the market. Our results show that the impact of stochastic correlation in modeling wrong way risk is significant because of the rich variability and heavier tails that our model produce when compared to constant correlation model. This way the stochastic correlation model may be able to capture the extreme events which are not possible to model by using constant correlation models. Tail risk should not be ignored when pricing any financial asset or a derivative. Thus, for a more conservative and risk-averse investor a stochastic correlation modeling approach can help her/him to hedge against the credit risk. \\

\appendix
\section{Deriving the value of the CCS contract}
\label{appendix:A}
Forward rates can be defined as the interest rates that can be locked today for an investment in a future time period. Forward rate can be understood through prototypical Forward Rate Agreement (FRA).
Forward Rate Agreement (FRA) is an interest rate derivative contract between two parties who want to protect themselves against future movements in interest rates. FRA can be defined by three time instants: current time t, the expiry time $T_{1}$, and the maturity time $T_{2}$. This contract gives its holder an interest-rate payment for the period between $T_{1}$ and $T_{2}$. At the maturity $T_{2}$, a payment based on the fixed rate K is exchanged against a floating payment based on the spot rate $L(T_{1}, T_{2})$. In other words, this contract let one to lock-in the interest rate between times $T_{1}$ and $T_{2}$ at a desired value K, for a contract with simply compounded rates. Therefore, the expected cash flows are discounted from time $T_{2}$ and $T_{1}$. At time $T_{2}$ one receives $\Delta(T_{1},T_{2})$KN units of cash and pays the amount $\Delta(T_{1},T_{2})L(T_{1}, T_{2})N$. Here N denotes the contract's nominal value and $\Delta(T_{1},T_{2})$ denotes the year fraction for the contract period $[T_{1}, T_{2}]$. The value of the FRA contract, at time $T_{2}$ can be expressed as

\begin{equation}
N . \Delta(T_{1},T_{2}) . (K - L(T_{1}, T_{2}))
\end{equation}

Furthermore, $L(T_{1}, T_{2})$ can also be expressed as 

\begin{equation}
 L(T_{1}, T_{2}) =  \frac{1 - P(T_{1}, T_{2})}{\Delta(T_{1},T_{2})P(T_{1}, T_{2}) }
\end{equation}

Substituting it into equation (8) 

\begin{equation}
N\Bigg[\Delta(T_{1},T_{2})K - \frac{1}{P(T_{1}, T_{2})} +1\Bigg]
\end{equation}

As per the no arbitrage theory, the implied forward rate between the time t and $T_{2}$ can be derived from two consecutive zero coupon bonds due to the equality (Filipovic, 2009)

\begin{equation}
P(t, T_{2}) = P(t, T_{1}) P(T_{1}, T_{2})
\end{equation}

The value of FRA in terms of simply compounded forward interest rate gives

\begin{equation}
FRA(t, T_{1}, T_{2}, \Delta(T_{1},T_{2}), N, K) = N P(t, T_{2}) \Delta(T_{1},T_{2})(K - F_{s}(t; T_{1}, T_{2})), 
\end{equation}
where $F_{s}(t; T_{1}, T_{2})$ is the simply compounded forward interest rate prevailing at time t for the expiry $T_{1} >t$ at maturity $T_{2} > T_{1}$, defined as

\begin{equation}
F_{s}(t; T_{1}, T_{2}) = \frac{P(t, T_{1}) - P(t, T_{2})}{\Delta(t, T_{2})P(t, T_{2}) } = \frac{1}{\Delta(T_{1},T_{2})} \Bigg(\frac{P(t, T_{1})}{P(t, T_{2})} -1\Bigg)
\end{equation}\\

Seeing CCS contract from X's side, as a portfolio of FRA, every individual FRA can be evaluated by using equation (A5). The the value of the CCS for X denoted $\pi_{payer}(t)$, is given as

\begin{equation}
\pi_{payer}(t)  =  \sum_{i = \alpha+1}^{\beta}FRA(t, T_{1}, T_{2}, \Delta(T_{i-1},T_{i}), N, F_{d})
\end{equation}

\begin{equation}
\pi_{payer}(t) = N\sum_{i = \alpha+1}^{\beta} \Delta_{i} P(t, T_{i})(F_{f}(t; T_{i-1}, T_{i})- F_{d}(t; T_{i-1}, T_{i}))
\end{equation}

Using equation (A5) and (A6) in (A8) we get 

\begin{equation}
\pi_{payer}(t)  =  N\sum_{i = \alpha+1}^{\beta} \Big[P^{f}(t,T_{i-1})  - P^{f}(t,T_{i}) - (P^{d}(t, T_{i-1}) - P^{d}(t,T_{i})\Big]
\end{equation}

The above sum can be decomposed into 2 sums
\begin{equation}
N\sum_{i = \alpha+1}^{\beta} (P^{f}(t, T_{i-1})  - P^{f}(t, T_{i})) + N\sum_{i = \alpha+1}^{\beta} (P^{d}(t, T_{i})  - P^{d}(t, T_{i-1}))
\end{equation}

The first sum can be simplified into 
\begin{equation}
N\sum_{i = \alpha+1}^{\beta} (P^{f}(t, T_{i-1})  - P^{f}(t, T_{i})) = N (P^{f}(t, T_{\alpha}) -  P^{f}(t, T_{\beta}))
\end{equation}

The second sum can be simplified into 
\begin{equation}
N\sum_{i = \alpha+1}^{\beta} (P^{d}(t, T_{i-1})  - P^{d}(t, T_{i})) = N (P^{d}(t, T_{\beta}) -  P^{d}(t, T_{\alpha}))
\end{equation}

However, the floating leg need to converted into domestic currency by multiplying it by the spot foreign exchange rate $\psi_{t}$ and the notionals are also exchanged at the maturity date. Using the relation $ N^{d}  = \psi_{0}N^{f}$ and the equations (A11) and (A12) we get 

\begin{equation}
\pi_{payer}(t) = N^{f}[\psi(t)(P^{f}(t, T_{\alpha}) - P^{f}(t, T_{\beta})] + N^{f}\psi(\mathbf{T}) - N^{d}[P(t,T_{\alpha}) - P(t, T_{\beta})] - N^{d},
\end{equation}
which is equation \eqref{exp2}.

\section{Data Description}
\label{appendix:B}

\renewcommand\arraystretch{2.5}
\begin{table}[h]
\centering
\tbl{Data Description }
{\begin{tabular}{|l|l|l|}
\hline
Data Type \ & Source \vspace{-0.08cm} \ & Time Period  \vspace{-0.08cm} \\
\hline
USD-LIBOR and EUR-LIBOR & \url{https://www.global-rates.com/}  & 1-day, 1-week, 3-months, 6-months and 1-year\\
\hline
USD-SWAP and EUR-SWAP  & \url{http://www.interestrateswapstoday.com/} & 1-10 years \\
\hline
USD YIELD & \url{https://www.feddata.com/}  & February 25, 2018 - February 03, 2019 \\
\hline
EUR YIELD  & \url{https://www.ecb.europa.eu/home/html/index.en.html} & February 25, 2018 - February 03, 2019 \\
\hline
EUR/USD FX  & \url{https://www.investing.com/} & January 05, 2014 - April 01, 2018.\\
\hline 
CDS & Bloomberg  & January 05, 2014 - April 01, 2018 \\
\hline
SWAPTION-USD & Bloomberg &  -\\
\hline
SWAPTION-EUR & Bloomberg & - \\
\hline
IMPLIED VOL of EUR/USD FX OPTION & Bloomberg  & - \\
\hline
\end{tabular}
\label{data-table}}
\end{table}
\end{document}